\documentclass[preprint,showpacs,amsmath,amssymb]{revtex4}
\usepackage{graphicx,amsmath,amssymb,bbm,times}
\usepackage{psfig}
\begin{document}
\title{Demonstration of a bright and compact source of tripartite
nonclassical light}
\author{Alessia Allevi}
\affiliation{C.N.I.S.M., U.d.R. Como, I-22100, Como, Italy}
\author{Maria Bondani} \email{maria.bondani@uninsubria.it}
\affiliation{National Laboratory for Ultrafast and Ultraintense
Optical Science - C.N.R.-I.N.F.M.\\ and C.N.I.S.M., U.d.R. Como, I-22100, Como, Italy}
\author{Matteo G. A. Paris}
\affiliation{Dipartimento di Fisica, Universit\`a degli Studi di Milano\\ and C.N.I.S.M., U.d.R. Milano, I-20133 Milano, Italy\\
and ISI Foundation, I-10133 Torino, Italy}
\author{Alessandra Andreoni}
\affiliation{Dipartimento di Fisica e
Matematica, Universit\`a degli Studi dell'Insubria\\ and C.N.I.S.M., U.d.R. Como, I-22100, Como, Italy}
\date{\today}
\begin{abstract}
We experimentally demonstrate the nonclassical photon number
correlations expected in tripartite continuous variable states obtained
by parametric processes. Our scheme involves a single nonlinear crystal,
where two interlinked parametric interactions take place simultaneously,
and represents a bright and compact source of a sub-shot-noise tripartite
light field. We analyze the effects of the pump intensities
on the numbers of detected photons and on the amount of noise reduction in some details, thus demonstrating a good agreement between the experimental data and a single-mode theoretical description.
\end{abstract}
\maketitle
\section{Introduction}
Multimode light beams endowed with nonclassical correlations, as those
exhibited by multipartite entangled states, represent a resource for
quantum technology. They are at the heart of enhanced quantum imaging,
either ghost imaging or ghost diffraction \cite{Dangelo2005, Gatti2008},
and represent a building block for the development of an integrated
quantum network.  In turn, nonlinear interactions involving multimode
beams of radiation have attracted much attention in the recent years,
either to realize all-optical information processing \cite{Kartaloglu}
or to generate nonclassical states of light \cite{Zhang}. Several
experimental schemes to generate multimode entangled states have been
suggested and demonstrated. The first example is provided by the
original continuous variable (CV) teleportation experiments in
Ref.~\cite{furu}, where one mode of a twin beam was mixed with a
coherent state, although no specific analysis was made on the
entanglement properties besides the verification of teleportation. A
similar scheme, where one mode of a twin beam is mixed with the vacuum,
has been demonstrated and applied to controlled dense coding
\cite{jing}. Moreover, a fully inseparable three-mode entangled state
has been generated and verified by mixing three independent squeezed
vacuum states in a network of beam splitters \cite{aoki}. Recently we
suggested and demonstrated a compact scheme to realize three-mode
entanglement by means of two interlinked $\chi^{(2)}$ interactions
occurring in a single nonlinear crystal in a type-I non-collinear
phase-matching geometry \cite{OLn,OEmoon}. Other schemes involving
cascaded interactions have been also analyzed either in periodically
poled crystals \cite{Rodionov} or in second-order nonlinear ones
\cite{Bradley,Olsencasc,Pfister}. Notice, however, that the use of a
single nonlinear medium makes the system more compact and robust
compared to the other schemes that have been suggested and demonstrated so far, in which
additional parametric sources and linear devices, such as beam
splitters, introduce unavoidable losses. Finally, parametric oscillators
have been suggested as a source of tripartite signal-idler-pump
entanglement in triply resonant cavities \cite{vil06}.
\par
In this paper we experimentally demonstrate the nonclassical photon
correlations exhibited by tripartite states generated by a single
nonlinear crystal, where two interlinked parametric interactions take
place simultaneously. Our scheme realizes a bright and compact source of
sub-shot-noise three-mode light beams and allows the implementation of
simultaneous ghost imaging and ghost diffraction protocols with enhanced
sensitivity.
\par
The paper is structured as follows: in the next Section we provide a
theoretical description of our system and evaluate correlations and
noise reduction as a function of the coupling parameters. In Section
\ref{sec:exp} we describe our experimental apparatus, illustrate the
results with focus on nonclassical photon-number correlations, and analyze
the sources of noise in some details. Section \ref{sec:out} closes the
paper with some remarks.
\section{Theoretical description}
\label{sec:theory}
In our scheme two interlinked interactions, namely a spontaneous
parametric downconversion process and a sum-frequency
generation, take place simultaneously in a single nonlinear
crystal. In principle, five modes $a_j$ are involved in the interactions, two
of which, say $a_4$ and $a_5$, are non-evolving undepleted
pumps and thus are included in the coupling coefficients (parametric approximation). The effective Hamiltonian
describing the interaction is thus given by
\begin{equation}  \label{intH2} H_{\mathrm{int}} = g_1
a_1^\dag a^{\dag}_3 + g_2 a_2^{\dag} a_3 + h.c.\;,
\end{equation}
where $g_1$ and $g_2$ are coupling coefficients linearly dependent on the pump fields $a_4$ and $a_5$, respectively. The
earliest studies on the dynamics and the quantum properties of the
states realized via this Hamiltonian can be traced back to the
works in Refs.~\cite{ea1,ea2}. The relevance of studying the
dynamics generated by the above Hamiltonian in details lies in the
fact that $H_{\textrm{int}}$ can be realized in a variety of different contexts,
from quantum optics \cite{Rodionov,Olsencasc,pap1,pap2,Guo} to
condensate physics \cite{PCB03,CPP04}. The coupling between
two optical modes and one vibrational
mode of a macroscopic object, such as a mirror, has been
considered \cite{PMV+03} and also ions trapped in a cavity
have been demonstrated to realize the Hamiltonian in
Eq.~(\ref{intH2}) for a suitable configuration \cite{LWH05}.
\par
The Hamiltonian admits the constant of motion
$\Delta(t) \equiv N_1(t) - N_2(t) - N_3(t) \equiv \Delta(0)$.
If we take the vacuum $|{\bf 0\rangle}\equiv |0\rangle_1 \otimes
|0\rangle_2 \otimes |0\rangle_3$ as the initial state, we have
$N_1(t)=N_2(t)+N_3(t)$
$\forall t$, being $N_{\emph{j}}(t)=\langle a^\dag_{\emph{j}}(t)
a_{\emph{j}}(t)\rangle$ the mean number of photons in the $j$-th
mode. Under these hypotheses the evolved state $|{\bf T}\rangle
=\exp\{-i H_{\rm int} t\}|{\bf 0}\rangle$ may  be written as
\begin{equation}
|{\bf T}\rangle =
\sum_{mr} \frac{N_2^{m/2}N_3^{r/2}}{(1+N_1)^{(1+m+r)/2}}
\sqrt{\frac{(m+r)!}{m! r!}}\:
|m+r,m,r\rangle \label{state}\;,
\end{equation}
where we omitted the time dependence of $N_{\emph{j}}$.
As a matter of fact the state in Eq.~(\ref{state}) is a
fully inseparable three-mode Gaussian state \cite{Allevi06}, \textit{i.e.}
a state that is inseparable with respect to any grouping of the
modes, thus permitting realizations of truly tripartite quantum
protocols such as conditional twin-beam generation and telecloning
\cite{pap1,pap2}. The mean numbers of photons $N_{\emph{j}}$ that appear in
Eq.~(\ref{state}) can be obtained by the Heisenberg evolution of
the field operators. In particular, by introducing $\Omega =
\sqrt{|g_2|^2-|g_1|^2}$ we have $N_1 = N_2+N_3$
and
\begin{eqnarray}\label{Ndit}
N_2 = \frac{|g_1|^2 |g_2|^2}{\Omega ^4}
\left[\cos{\Omega t}-1 \right]^2 \quad
N_3 = \frac{|g_1|^2}{\Omega ^2} \sin^2(\Omega t) \;.
\end{eqnarray}
We see that when $|g_2|^2 >|g_1|^2$ the dynamics is
oscillatory; viceversa, when $|g_1|^2
>|g_2|^2$ we find an exponential behavior.
\par
The above description of the system has been derived under the
hypothesis of perfect frequency-matching and phase-matching conditions among single-mode fields and the time $t$ appearing in Eqs.~(\ref{Ndit}) represents
the interaction time inside the crystal.
In this case we did not need to take into account the existence of
temporal modes and spatial coherence areas. On the other hand, if the
pump fields are pulsed, the generated fields are
temporally multimode \cite{Paleari}. Moreover, in a non-collinear interaction geometry, the momentum conservation in the transverse direction can be satisfied in more than one way. Thus coherence areas exist \cite{Joobeur,Allevi06}, whose angles of divergence depend on several
parameters, such as the pumps intensities, the distance from the
collinear interaction geometry and the wavelengths of the generated
fields.
It is interesting to point out that in the CV regime
the demonstration of the entangled nature of the state in
Eq.~(\ref{state}) critically depends on the correct collection of these
coherence areas~\cite{subshot}. In fact, collecting light from more than a
single coherence area corresponds to the introduction of spurious light,
while collecting less than a coherence area determines a loss of
information, which is detrimental to the investigations of the
nonclassical properties. In addition, we have to select a triplet of
areas as there is a one-to-one correspondence between the coherence
areas in each field. To achieve such a selection we can apply a criterion which represents a necessary but not sufficient condition, based on the study of the correlation in the number of photons.  In fact, due to the constant of motion, the state in Eq.~(\ref{state}) is endowed with perfect correlations in the number of photons. The three-mode photon-number distribution is given by
\begin{equation}
P_T(n,m,r) = \delta_{n,m+r} \frac{N_2^m N_3^r}{(1+N_1)^{1+m+r}}
\frac{(m+r)!}{m! r!}
\label{phnstat}\;,
\end{equation}
from which we can derive the photon-number correlation
coefficients between the components of the entangled state. In
particular, due to the conservation law,
we expect the existence of strong intensity correlations between the
number of photon $n_1$ and the sum of the other two, say $n_2 +
n_3$.  In order to quantify correlations we denote
by $\gamma(n_j,n_k) =\langle n_j n_k \rangle -
\langle n_j\rangle \langle n_k \rangle$ and $\sigma^2 (n_j)
= \langle {n_j}^2 \rangle - \langle n_j \rangle^2$ the covariance
and the variance of the number of photons, respectively, and introduce the correlation
coefficients as follows
\begin{equation}
\epsilon_{j,k} =
\frac{\gamma(n_j,n_k)}{\sigma(n_j)\sigma(n_k)}\:.
\end{equation}
Upon exploiting Eq.~(\ref{phnstat}) we have that the correlation
coefficient $\epsilon_{1,2+3}$ is identically equal to one,
independently of the number of photons generated by the interlinked
interactions.
On the other hand, for the partial photon-number correlations
we obtain expressions that do depend on the mean
number of photons involved. Upon writing $N_k = \beta_k N$ where
$\beta_1 = \beta_2 + \beta_3$ and $N$ is the total number of photons
of the state we have
\begin{align}\label{corrpart}
\epsilon_{1,k} &= \sqrt{
\frac{N_k( 1+N_1)}{N_1(1+N_k)}} \stackrel{N\gg
1}{\simeq}1 - \frac{\beta_1-\beta_k}{2 \beta_1 \beta_k N} \\
\epsilon_{2,3} &= \sqrt{\frac{N_2
N_3}{(1+N_2)(1+N_3)}} \stackrel{N \gg 1}{\simeq} 1 -
\frac{\beta_2+\beta_3}{2 \beta_2 \beta_3 N}
\end{align}
where from now on $k=2,3$.
As the detectors we used to perform the correlation measurements
are not ideal, we have to rewrite the expressions of the correlation
coefficients by taking into account
the non-unit quantum efficiency of the detection apparatus.
The probability operator-valued measure
(POVM) of each detector, describing the statistics of detected
photons, is given by a Bernoullian convolution of the ideal
number operator spectral measure
\begin{equation}
\hat{\Pi}_{m_j}={\eta_j}^{m_j}\sum_{n_j=m_j}^\infty
(1-\eta_j)^{n_j-m_j} \left( \begin{array}{c}  n_j\\ m_j
\end{array}\right)\: |n_j\rangle \langle n_j| \label{POVM}
\end{equation}
with $j=1,2,3$.  Equation~(\ref{POVM}) can be exploited to calculate
the expressions of mean number, $\langle m_j \rangle$, and variance, $\sigma^2(m_j)$,
of the detected photons $m_j$ in terms of the mean number of the photons $n_j$
and of its variance $\sigma^2(n_j)$ \cite{jointdiff}
\begin{eqnarray} \label{moments}
M_j \equiv \langle m_j \rangle &=& \eta_j \langle n_j\rangle = \eta_j N_j\\
\sigma^2(m_j) &=& \eta_j^2 \sigma^2(n_j) + \eta_j(1-\eta_j)
N_j \nonumber
\end{eqnarray}
We notice that, in general, the statistical distribution of the
number of detected photons is different from that of the number of
photons. Nevertheless, the correlation coefficients, $\epsilon^{m}$, calculated for
the detected photons can also assume high values; in particular,
the correlation coefficient calculated between $m_1$ and the sum
$m_2+m_3$ reads as follows
\begin{equation}
\epsilon_{1,2+3}^{m}=\frac{\eta (1+N_1)}{(1+\eta N_1)}\stackrel{N
\gg 1}{\simeq}1 - \frac{1-\eta}{\eta}\frac{1}{\beta_1 N} \label{epsdetph}
\end{equation}
where we have assumed that all the detectors have the same quantum
efficiency $\eta$. In turn, the partial correlations are given by
\begin{align}\label{corrpartm}
\epsilon_{1,k}^m & \stackrel{N\gg
1}{\simeq}1 - \frac{\beta_1+\beta_k - 2 \eta \beta_k }{2 \beta_1 \beta_k N}\\
\epsilon_{2,3}^m & \stackrel{N \gg 1}{\simeq} 1 -
\frac{\beta_2+\beta_3}{2 \eta \beta_2 \beta_3 N}\:,
\end{align}
and approach unit value for large $N$ values.
\par
As a matter of fact a large value of the correlation indices is
not sufficient to discriminate between quantum and classical
correlations \cite{jointdiff}. A trivial example is given by the
mixture $\varrho= \sum_{nmr} P_T (n,m,r)
|n\rangle\langle n| \otimes
|m\rangle\langle m| \otimes
|r\rangle\langle r| $, which, with $P_T (n,m,r)$ given as in Eq.~(\ref{phnstat}),
exhibits the same correlations of the state $|{\mathbf{T}}\rangle$. A more realistic example is provided by
the tripartite state generated by sending a
thermal state on two subsequent beam-splitters, whose second port is
unexcited: the state is classical and shows large intensity correlations,
approaching unit value for large mean photon numbers \cite{CEWQO}.
\par
In order to obtain a proper marker of nonclassicality we may take into
account the difference photocurrents $d_{j,k}=m_j-m_k$ \cite{subshot}
and build the so-called noise reduction factor
\begin{align}
R_{j,k} = \frac{\sigma^2(d_{j,k})}{\langle m_j\rangle + \langle m_k\rangle}
\:,
\end{align}
which is smaller than one for nonclassically correlated states.  Note
also that, for states generated by the Hamiltonian in Eq.(\ref{intH2}),
the existence of sub-shot noise photon-number correlations is a
sufficient condition for entanglement, \textit{i.e.} the condition of negative
partial transpose is subsumed by the condition of sub-shot noise
correlations \cite{pap2}.  By using Eqs.~(\ref{moments}) we may write
\begin{align}
R_{j,k} = 1-\eta +
\frac{\eta\left[\sigma^2(n_j)+\sigma^2(n_k)-2\gamma(n_j,n_k)\right]}{\langle
n_j\rangle + \langle n_k\rangle}
\:,
\end{align}
for the noise reduction of bipartite correlations whereas, for
the difference photocurrent between the mode $a_1$ and the sum
of the other two modes, we have $ R \equiv R_{1,2+3} $
\begin{align}
R & = 1- \eta + \frac{\eta
\left[\sum_p \sigma^2(n_p)+ 2\Gamma(\{n_k\})
\right]}{\sum_p \langle n_p \rangle}
\label{NRF}
\end{align}
where
\begin{align}
\Gamma(\{n_k\}) & =  \gamma(n_2,n_3) - \gamma(n_1,n_2) -
\gamma(n_1,n_3)
\:.
\end{align}
For the state in Eq.~(\ref{state}) we have
\begin{equation}R=1-\eta\:,\label{RT}\end{equation}
which shows that state $|\mathbf{T}\rangle$ exhibits
nonclassical tripartite correlations for any value of the
mean number of photons. Besides, Eq. (\ref{RT}) says that
the noise reduction can be detected for any value of the
quantum efficiency $\eta$.
The corresponding bipartite quantities read as follows
\begin{align}
R_{1,k} & = 1 + \frac{\eta \left[(N_1-N_k)^2 - 2 N_k\right]}{N_1+N_k}
\stackrel{N\gg 1}{\simeq} \eta N
\frac{(\beta_1-\beta_k)^2}{\beta_1+\beta_k}\label{bipR}
\\
R_{2,3} & = 1 + \frac{\eta (N_2-N_3)^2}{N_2+N_3}
\stackrel{N\gg 1}{\simeq} \eta N
\frac{(\beta_2-\beta_3)^2}{\beta_2+\beta_3}\label{Rpar}\:,
\end{align}
and say that the correlations between modes $a_2$ and $a_3$ are
always classical whereas the correlations between mode $a_1$ and
either mode $a_2$ or mode $a_3$ may be nonclassical in certain regimes.
More specifically, we have $R_{1,k}<1$ if $N_1< N_k + \sqrt{2 N_k}$.
Since $N_1=N_2+N_3$ we may have both the noise reduction parameters below
the classical threshold only for an overall energy of the
state $N_1+N_2+N_3 < 4$.
\section{Experiment}
\label{sec:exp}
\begin{figure}
\begin{center}
\resizebox{0.8\columnwidth}{!}{%
 \includegraphics[angle=270,width=1\textwidth]{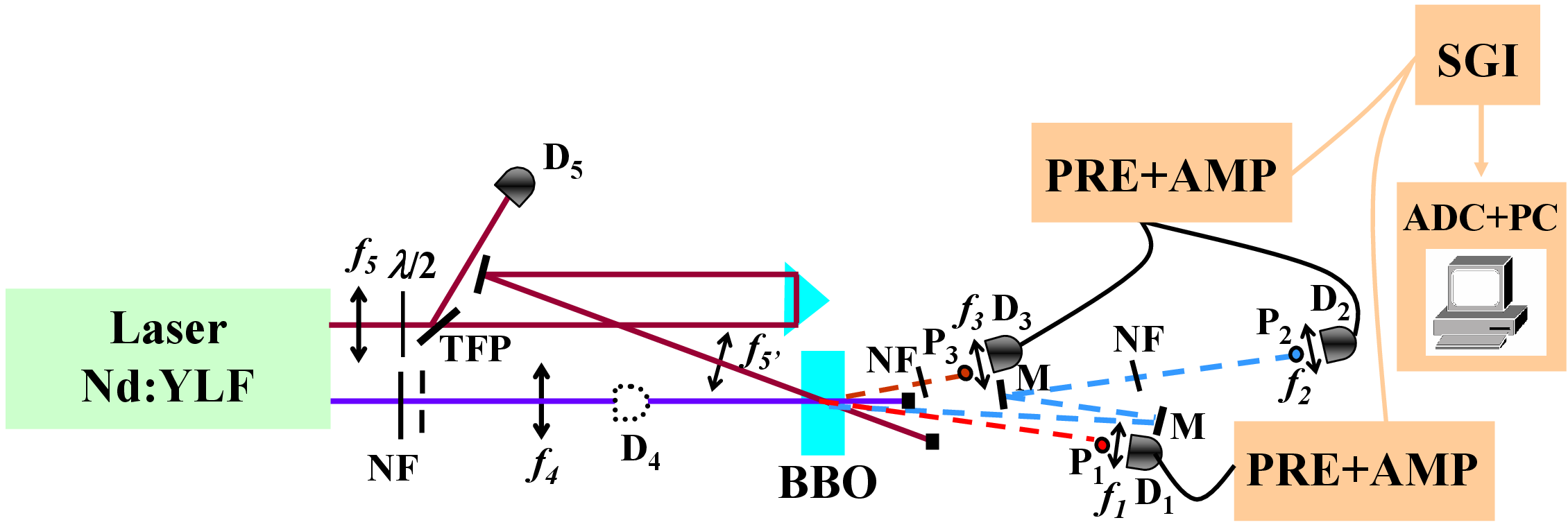}}
 \caption{Scheme of the experimental setup: BBO, nonlinear crystal;
NF, variable neutral-density filter; $\lambda /2$, half-wave
plate; TFP, thin-film plate polarizer; P$_{1-3}$, pin-holes;
$f_{1-5,5'}$, lenses; D$_{1-3}$, p-i-n photodiodes; D$_{4,5}$,
thermal detectors; M, Aluminum mirrors; PRE+AMP, low-noise
charge-sensitive pre-amplifiers followed by amplifiers; SGI,
synchronous gated-integrator; ADC+PC, computer integrated
digitizer.} \label{setup}
\end{center}
\end{figure}

\begin{figure}
\begin{center}
\resizebox{0.8\columnwidth}{!}{%
 \includegraphics[angle=270,width=1\textwidth]{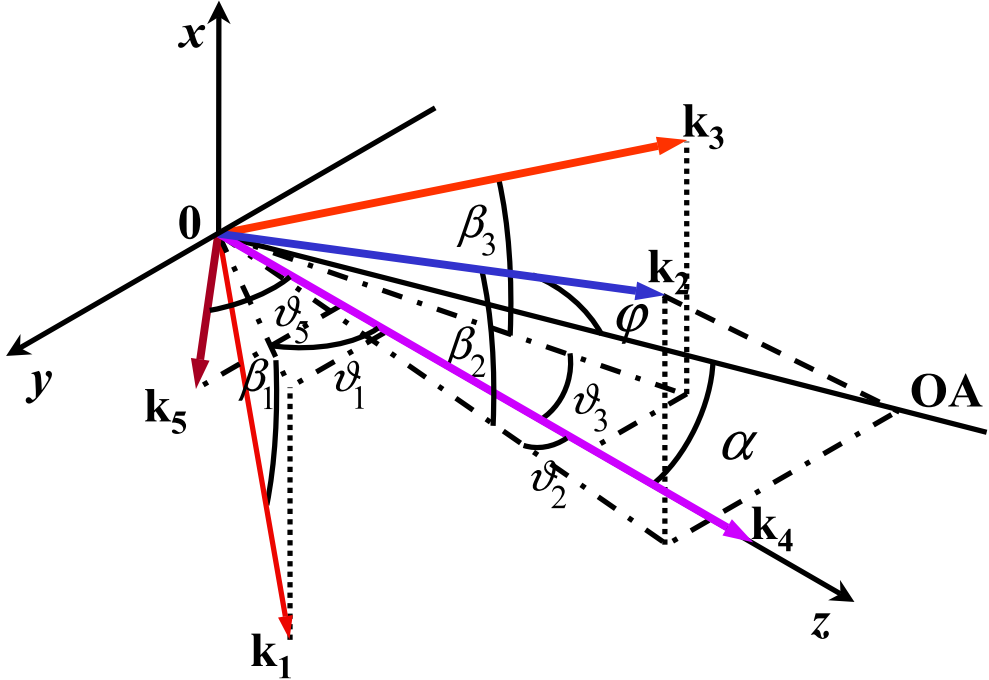} }
\caption{Scheme of the phase-matched interlinked interactions:
$(x,y)$-plane coincides with the crystal entrance face; $\alpha$,
tuning angle; $\beta_{\emph{j}}$'s, angles to $(y,z)$-plane;
$\vartheta_{\emph{j}}$'s, angles on the $(y,z)$-plane; $\varphi$,
angle to the optical axis (OA).} \label{f:schemeTRIBIT}
\end{center}
\end{figure}

The experimental scheme used to generate the nonclassical
state of Eq.~(\ref{state}) is depicted in
Fig.~\ref{setup}.\\
The harmonics of a continuous-wave
mode-locked Nd:YLF laser regeneratively amplified at a repetition
rate of 500 Hz (High Q Laser Production, Hohenems, Austria)
provide the two pump fields. In particular, the third harmonic
pulse at 349 nm ($\sim 4.45$ ps pulse-duration) is exploited as
the pump field $a_4$ in the downconversion process, whereas
the fundamental pulse at 1047 nm ($\sim 7.7$ ps pulse-duration) is
used as the pump field $a_5$ in the upconversion process. The two
processes must simultaneously satisfy energy-matching
($\omega_{4}=\omega_{1}+\omega_{3}$,
$\omega_{2}=\omega_{3}+\omega_{5}$) and phase-matching
(${\mathbf k^e_4}={\mathbf k^o_1}+{\mathbf k^o_3}$, ${\mathbf
k^e_2}={\mathbf k^o_3}+{\mathbf k^o_5}$) conditions, in which
$\omega_{j}$ are the angular frequencies, ${\mathbf
k}_j$ are the wavevectors and suffixes $o,e$ indicate
ordinary and extraordinary field polarizations.
As depicted in Fig.~\ref{f:schemeTRIBIT}, we set the pump-field $a_4$
direction so that the wavevector ${\mathbf k_4}$ is normal to
the crystal entrance face and propagates along the $z$-axis of the
medium. We also align the wavevector ${\mathbf k_5}$ of the
other pump field $a_5$ in the plane
($y$, $z$) containing the optical axis (OA) of the crystal and
the wavevector ${\mathbf k_4}$.
The nonlinear medium is a $\beta$-BaB$_2$O$_4$ crystal
(BBO, Fujian Castech Crystals, China, 10 mm $\times$ 10 mm cross
section, 4 mm thickness) cut for type-I interaction
($\vartheta_{\mathrm{cut}} = 38.4$ deg), into which both pumps are strongly focused. Typical intensity values of the pumps were $\sim 5$ GW/cm$^2$ for $a_4$ and $\sim 2$ GW/cm$^2$ for $a_5$. The required superposition in time of
the two pumps is obtained by a variable delay line.\\
With reference to Fig.~\ref{f:schemeTRIBIT}, we indicate as
$\vartheta_j$ the angles in the plane ($y$, $z$) formed by each
wavevector with ${\mathbf k_4}$ and as $\beta_j$ the angles of each
wavevector with respect to this plane. For the experimental
realization of the interaction scheme we choose the solutions in the
plane ($y$, $z$), thus $\beta_j = 0$ for $j=1-3$: in particular, we
sent the pump field $a_5$ at an
external angle $\vartheta_{5,ext}=-24.47$ deg with respect to the
other pump field $a_4$. Under these hypotheses, for $\lambda_{1} =
632.8$ nm, $\lambda_{2} = 446.4$ nm and $\lambda_{3} = 778.2$ nm,
we calculated the following external interaction angles with respect
to the pump field $a_4$: $\vartheta_{1,ext}=-9.78$ deg,
$\vartheta_{2,ext}=-3.25$ deg and $\vartheta_{3,ext}=+12.06$ deg \cite{OEmoon}.
\par
The preliminary use of a He:Ne laser as the seed field allowed us to
position three pin-holes on the path of the three generated fields in
such a way that then, when operating the system from vacuum (\textit{i.e.} in
the absence of any seed fields), we could collect a triplet of
coherence areas.  Distances and sizes of the pin-holes were chosen by
searching for the condition of maximum intensity correlations between
the generated fields \cite{CEWQO}. In fact, as shown in
Section~\ref{sec:theory}, we expect strong intensity correlations not
only between the number of detected photons $m_1$ and the sum of the
other two, but also between $m_1$ and $m_2$, $m_2$ and $m_3$ and $m_1$
and $m_3$. By applying this criterion, we finally decided to put two
pin-holes of 30 $\mu$m diameter at distances $d_1 = 60$ cm and $d_3 =
49$ cm from the BBO along the path of the signal beam at 632.8
nm and of the idler beam at 778.2 nm, respectively. The two different
distances were chosen to compensate for the difference in the divergence
of signal and idler due to their wavelengths \cite{Allevi06}. Moreover,
as the beam at 446.4 nm has a divergence smaller than those of the other
two fields, we selected it by means of a 50 $\mu$m diameter pin-hole
placed at a distance $d_2 = 141.5$ cm from the crystal.
\par
The light, suitably filtered by means of bandpass filters located in
front of each pin-hole, was focused on each detector by a lens
($f_1=f_3=25$~mm, $f_2=10$~mm). Since we performed measurements in the
macroscopic intensity regime (more than 1000 photons per coherence area), we used
three p-i-n photodiodes (two, D$_{1,2}$ in Fig.~\ref{setup}, S5973-02
and one, D$_3$, S3883, Hamamatsu, Japan) as the detectors. In order to
obtain the same overall detection efficiency (bandpass filter plus
detector) on the three arms, we put two adjustable neutral-density
filters in the pathways of $a_2$ and $a_3$, thus obtaining the same
value $\eta=0.28$ on the three arms. The current output of the detectors
was amplified by means of two low-noise charge-sensitive pre-amplifiers
(CR-110, Cremat, Watertown, MA) followed by two amplifiers (CR-200-4
$\mu$s, Cremat).
We connected the detectors $D_2$ and $D_3$ to the same amplifier device by means of a T-adapter. The two amplified outputs were then integrated by synchronous gated-integrators (SGI
in Fig.~\ref{setup}) operating in external trigger modality (SR250,
Stanford Research Systems, Palo Alto, CA). The voltage outputs were then
sampled, digitized by a 12-bit converter (AT-MIO-16E-1, DAQ National
Instruments) and recorded by a computer.
\par
In the following we discuss the measurements of the intensities of field $a_1$ and of the sum $a_2+a_3$ as, according to Eqs.~(\ref{RT})-(\ref{Rpar}), we expect a nonclassical behavior.
Partial measurements performed by alternatively blocking the light impinging on the detectors $D_2$ and $D_3$ were not very reliable as the numbers of detected photons on the two fields separately were too close to the electronic noise of the detection chain. This is an important drawback as, for all calculations that follow based on experimental data, we must take into account the electronic noise that we measured in the absence of light \cite{subshot}.
\par
As the pump fields are pulsed and their duration is longer than the
characteristic time of the nonlinear processes, the distributions of the
detected photons collected by the pin-holes are temporally multimode
\cite{Paleari}. The same is also true for the statistical distribution
of the sum $m_2+m_3$. Moreover these distributions should be
characterized by the same number of modes \cite{Allevi06}.
\par
From the experimental point of view, the main difficulty to be overcome was the correct selection of a triplet of coherence areas. In fact, in the CV domain we have to avoid spurious light that could be detrimental to the experimental results; moreover, the interaction scheme presented here involves not only two generated fields, but also a third one, which obviously makes the detection more critical. Finally, we have two pump fields instead of one and in particular we are not able to exactly measure the effective portions of them that interact into the crystal. In spite of all these difficulties, we characterized the state produced by the interlinked interactions and in particular we proved its quantum nature by performing sub-shot noise photon-number correlation measurements as a function of the pump fields intensities.
In fact, as remarked in Section \ref{sec:theory}, the evaluation
of the noise reduction factor $R$ for the distribution of the difference
photocurrent $d=m_1-(m_2+m_3)$ provides a sufficient condition in
order to test the quantum nature of the generated state.
\par
We firstly investigated the evolution of the mean number of photons as a function of the intensity of one of the two pumps by keeping fixed the intensity of the other one \cite{manuscript}. In fact, if on one hand this analysis allows us to verify that the mean number of photons
does not depend on the correct selection of the coherence areas, on the
other one it is essential for the determination of the effective values
of the pump fields intensities from the fitting curves.
\par
\begin{figure}
\begin{center}
\resizebox{1.0\columnwidth}{!}{%
 \includegraphics[width=1\textwidth]{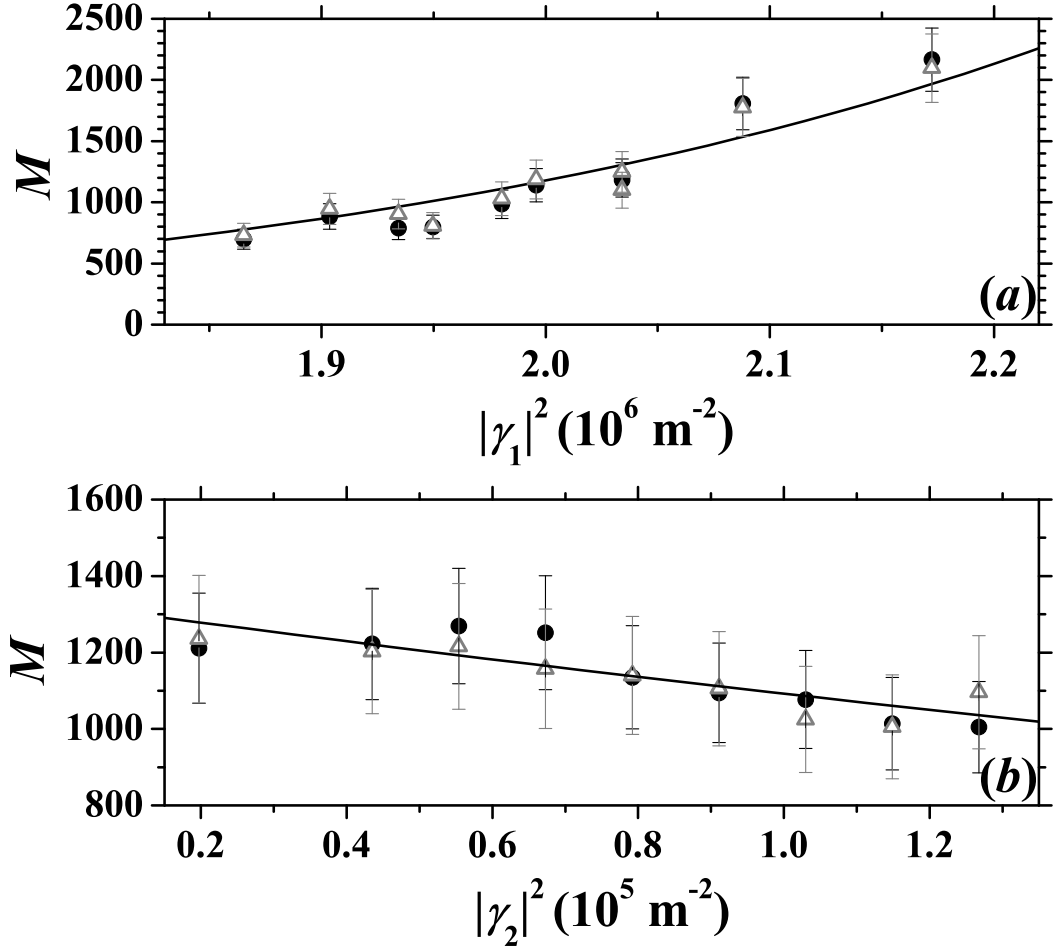} }
\caption{(a) Evolution of the mean numbers of detected photons as a
function of $|\gamma_1|^2$ which is proportional to the intensity of field $a_4$ for $|\gamma_2|^2=8.17\times10^5$ m$^{-2}$. Black circles:
measured values of $M_1$; grey triangles: measured values of $M_2+M_3$;
solid straight line: fitting curve. (b) Evolution of the mean numbers of
detected photons as a function of $|\gamma_2|^2$ which is proportional
to the intensity of field $a_5$ for $|\gamma_1|^2=1.52\times10^6$~m$^{-2}$. Black circles: measured values of $M_1$; grey triangles:
measured values of $M_2+M_3$; solid straight line: fitting curve.}
\label{andamento}
\end{center}
\end{figure}

As a first check we studied the evolution of the mean number of detected
photons, $M_1$ and $M_2+M_3$, as a function of the intensity of field
$a_4$ for a fixed value of the intensity of field $a_5$.
Note that temporal evolution in Eqs.~(\ref{Ndit})
is transformed into spatial evolution by identifying $\sqrt{|g_2|^2-|g_1|^2} t$ with
$\sqrt{|\gamma_2|^2-|\gamma_1|^2}z$, $z$ being the effective interaction length
\cite{dichro}. In
the experimental condition each $M_j$ represents the total mean number
of photons detected beyond each pin-hole; actually, it can be
expressed as $M_j=\mu \langle m_j \rangle$, where $\mu$ is the number of
temporal modes and $\langle m_j \rangle$ the average population of each
mode. To vary the intensity of field $a_4$, we changed its energy by
means of an adjustable neutral-density filter. For each energy value, measured by
means of a movable thermal detector ($D_4$ in Fig.~\ref{setup}, mod. 03A-P-CAL-SH, Ophir Optronics Ltd., Jerusalem, Israel), we measured the mean number of photons by averaging
over 50000 subsequent laser shots.
In Fig.~\ref{andamento}(a), we show the measured values of $M_1$ and
$M_2+M_3$ as functions of $|\gamma_1|^2$, for a fixed value of
$|\gamma_2|^2$. Note that $|\gamma_1|^2 \propto E_4/(\pi
r_4^2 \hbar \omega_4 \tau_4)$, $E_4$ being the pulse energy of field
$a_4$, $\tau_4$ the pulse duration and $r_4$ the beam radius. The
experimental data are displayed together with their common fitting curve,
obtained from Eqs.~(\ref{Ndit}) with $|\gamma_2|^2$ as the parameter and
$|\gamma_1|^2$ as the variable.  In this case we get
$|\gamma_2|^2=8.17\times10^5$ m$^{-2}$ and $|\gamma_1|^2$ in the range
$1.86\times10^6 - 2.17\times10^6$ m$^{-2}$. Note that, as expected,
the experimental data satisfy the photon-number conservation law as they
are almost superimposed. The best fitting curve has been obtained
allowing a slight difference in the quantum efficiency values of the
detection chains and finding the values from the conservation law. We
found $\eta_1 =0.31$ and $\eta_{sum} = 0.28$. The difference is within
the error justified by the tolerance of the pin-holes sizes (${\varnothing}_1={\varnothing}_3=30 \pm 2$ $\mu$m and ${\varnothing}_2=50 \pm 3$ $\mu$m) and it is also justified by
possible imperfections in the positioning of the pin-holes at the right
distances from the crystal.
\par
As a second check, we studied the evolution of $M_1$ and $M_2+M_3$ as a
function of the intensity of field $a_5$, by keeping the intensity of
field $a_4$ fixed. To change the energy of field $a_5$ we placed a
half-wave plate on the pathway of the infrared pump field. A movable
thin-film plate polarizer was used to measure the energy fraction
corresponding to the ordinarily polarized component of the field for
each step of rotation of the $\lambda/2$ plate. For each energy value,
measured by means of the thermal detector ($D_5$ in Fig.~\ref{setup}), we measured the mean number of photons by averaging over 50000 subsequent laser shots.
In Fig.~\ref{andamento}(b), we show the measured values of $M_1$ and
$M_2+M_3$ as functions of $|\gamma_2|^2$, for a fixed value of
$|\gamma_1|^2$. Also in this case, the experimental data are plotted
together with the fitting curve of the two sets of data obtained from Eqs.~(\ref{Ndit}).
Obviously, we have to interchange the roles of the pumps: in fact, here
$|\gamma_1|^2$ is treated as the parameter and $|\gamma_2|^2$ as the
variable. In particular, we obtained $|\gamma_1|^2=1.52\times10^6$
m$^{-2}$ and $|\gamma_2|^2$ in the range ($1.97\times10^4 -
1.27\times10^5$) m$^{-2}$.  Even in this case, the experimental data
satisfy the conservation law as they are almost superimposed and the optimization of the quantum efficiencies still gives very small
corrections: $\eta_1$ = 0.283 and $\eta_{sum}$ = 0.28.
\par
\begin{figure}
\begin{center}
\resizebox{1.0\columnwidth}{!}{%
\includegraphics[width=1\textwidth]{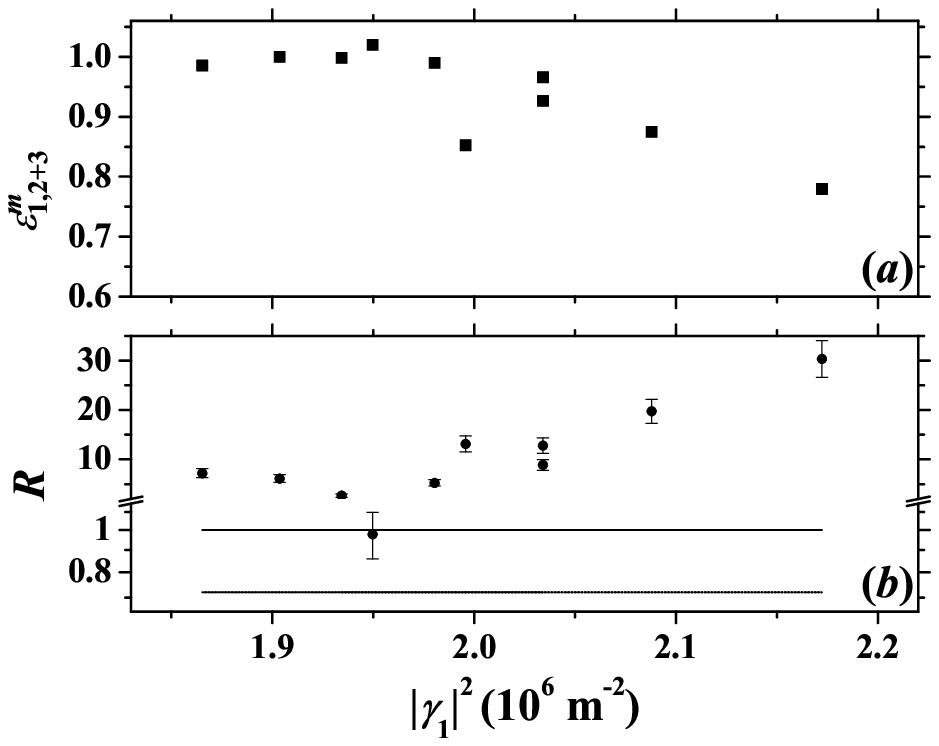}}
\caption{(a) Intensity correlation coefficient and (b) quantum
noise reduction $R$ (note the axis break) as functions of
$|\gamma_1|^2$ for $|\gamma_2|^2=8.17\times10^5$ m$^{-2}$.}
\label{corr&Ruv}
\end{center}
\end{figure}

By exploiting the values of the pump fields intensities obtained from the
fitting curves, we can investigate the behavior of the correlation
coefficient $\epsilon_{1,2+3}^{m}$ (see Eq.~(\ref{epsdetph})) and of the noise reduction $R$ (see Eq.~(\ref{RT})).
First of all, in Fig.~\ref{corr&Ruv} we show the intensity correlation
coefficient, (a), and the noise reduction, (b), as functions of
$|\gamma_1|^2$ by keeping fixed the value of $|\gamma_2|^2$. During
these measurements the collection areas were kept fixed (same pin-holes
located at the same distances as above). The variation of the
correlation coefficient as a function of the pump field intensity through
$|\gamma_1|^2$ is indeed not so strong, but the noise reduction
factor is critically dependent on the changes in the intensity value. In
fact, very much as in the case of the twin-beam state \cite{subshot}, there is
an optimum condition at which $R$ is minimum and, correspondingly, the
value of the correlation coefficient is maximum.
Note that $|\gamma_1|^2$ is larger than $|\gamma_2|^2$ in
the entire range of variation. Moreover, we note that increasing the
pump intensity, hence $|\gamma_1|^2$, also increases the size of the coherence
areas so that they are only partially transmitted by the pin-holes. On
the other hand, lowering the pump intensity reduces the size of the
coherence areas and allows uncorrelated light to pass the
pin-holes. Note that the values of $R$ corresponding to the selection of
more than a single coherence area remain quite close to the shot-noise
limit as the information contained in the area is not lost, but only
made more noisy. On the contrary, the selection of only a part of the
areas causes a loss of information that determines a more remarkable increase of $R$ above the shot-noise limit (note the axis break). This result represents an indication of the need of a perfect matching of the pin-hole areas in order to obtain sub-shot noise correlations.
\par
\begin{figure}
\begin{center}
\resizebox{1.0\columnwidth}{!}{%
 \includegraphics[width=1\textwidth]{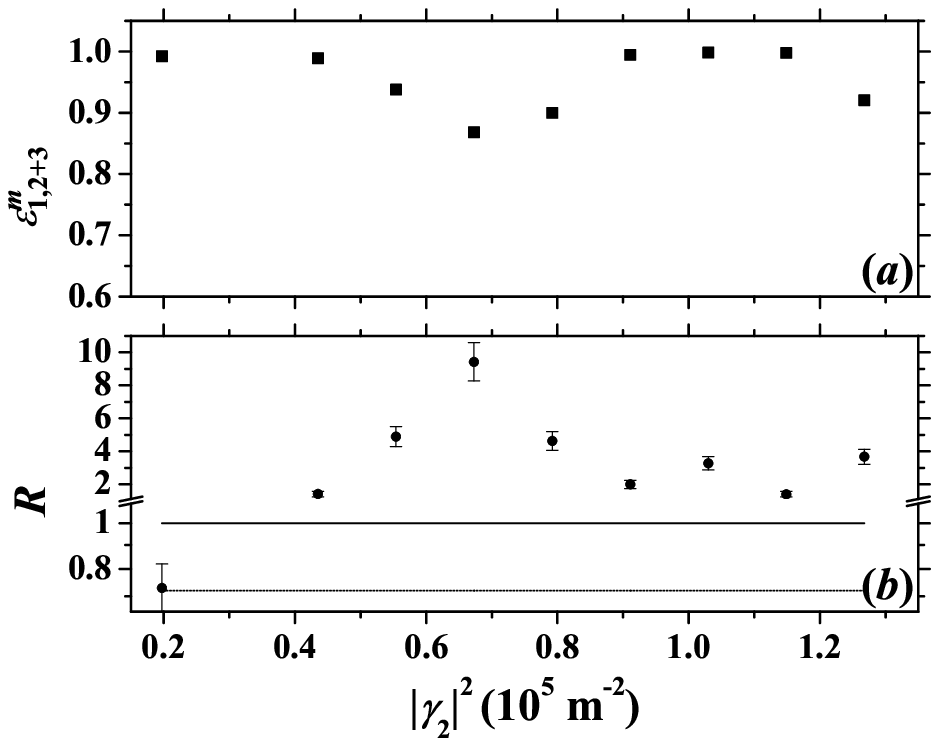} }
\caption{(a) Intensity correlation coefficient and (b) quantum
noise reduction $R$ (note the axis break) as functions of $|\gamma_2|^2$ for $|\gamma_1|^2=1.52\times10^6$ m$^{-2}$.}
\label{corr&Rir}
\end{center}
\end{figure}

Secondly, we investigated the intensity correlation coefficient and the
noise reduction as functions of $|\gamma_2|^2$ by keeping fixed the
value of $|\gamma_1|^2$. Also in this case the collection areas were
kept fixed by using the same pin-holes as before located at the same
distances. The intensity regime in which these measurements were
performed is different from the previous one as the absolute values of
the two pump fields intensities are smaller than in the other case.
However, $|\gamma_2|^2$ is again smaller than $|\gamma_1|^2$ in all its
range of variation. For all these reasons, the variations in the
experimental values of the correlation coefficient and of the noise
reduction are smaller (see Fig.~\ref{corr&Rir}). Moreover, the minimum
value of $R$ is quite near to the lower limit $R_{min}=0.72$ fixed by the quantum efficiency. In fact, the use of less intense pumps reduces the quantity of spurious light that can be revealed by the detectors; in addition, the fluctuations of the laser source and
the possible discrepancy of its
photon-number distribution with respect to the ideal Poissonian
statistics play a less important role \cite{jointdiff}.
\par
\begin{figure}
\begin{center}
\resizebox{1.0\columnwidth}{!}{%
 \includegraphics[width=1\textwidth]{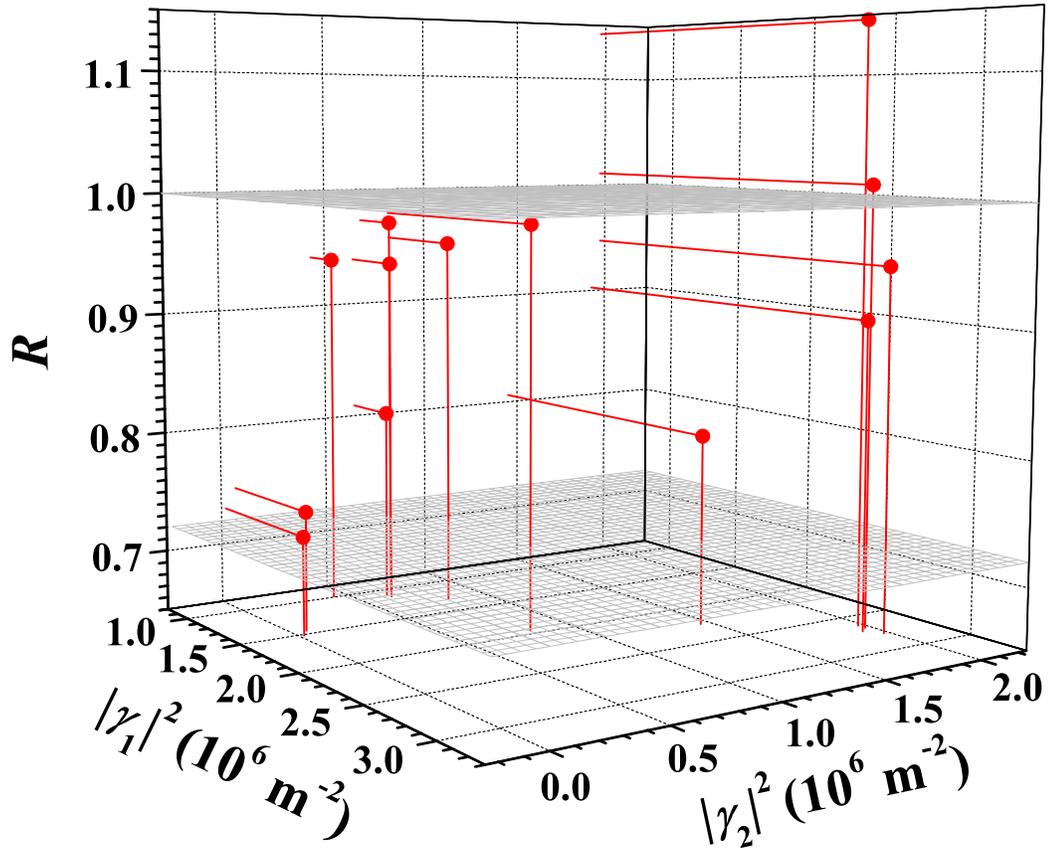} }
\caption{Noise reduction, $R$, as a function of
$|\gamma_1|^2$ and $|\gamma_2|^2$.}
\label{f:subshot}
\end{center}
\end{figure}

As a further investigation, we performed other measurements in order to
study how critical is the sub-shot noise condition with respect to a
slight change in the values of the intensities of the two pump fields. In
particular, we verified that it is always possible to choose the pumps
in such a way that only micro-metric adjustments of the pin-holes
positions are necessary to select the coherence areas. In
Fig.~\ref{f:subshot} we show a number of sub-shot noise measurements
obtained for different pairs of pump values. As we can see, not all the
measurements reach the optimum minimum value, $R \sim 0.72$, due to
residual imperfections in the selection of the coherence areas.
In particular, as we remarked above, this operation is more
critical when the intensity values are higher
because other noise sources become important \cite{jointdiff}. However, we want to emphasize that
there are many possible choices of the pumps values that allow us to
perform sub-shot noise measurements thus demonstrating that our scheme is
particularly versatile and useful for several applications in different
photon-number regimes.
\section{Conclusions and outlooks}
\label{sec:out}
In conclusion, we have presented the experimental realization of an entangled state that involves three modes of radiation in the
macroscopic regime. We verified the quantum nature of the state produced by our all-optical interaction scheme by means of sub-shot noise photon-number
correlations, which also subsumes the inseparability condition.
In particular, we investigated how critical is the
sub-shot noise condition by studying its dependence on the intensities of the two pump fields. In spite of the difficulties in measuring the light of  triplet coherence areas correctly and in avoiding the detection of spurious light, we obtained quite relevant results that could be further optimized. In the immediate future we plan to use three acquisition chains instead of only two to separately and simultaneously detect the three fields. Moreover, in order to reduce the noise that is detrimental to the shot-noise reduction factor we still intend to operate in the macroscopic regime, but with lower numbers of photons and to use hybrid photodetectors, which are endowed with a reasonable quantum efficiency ($\eta \simeq 0.4$) and a linear response in the mesoscopic regime (up to a few hundreds of detected photons), instead of
the p-i-n photodiodes. We also plan to modify our collection system by
using optical fibers in order to avoid spurious light and to minimize the uncertainty in the collection areas.  The experimental improvements would make the whole system more easily controllable and suitable for several applications, such as the
production of conditional twin-beam states and the generation of
quasi-Fock states with a number of photons sensibly greater than one.
Overall, our system represents a robust and tunable scheme
to obtain nonclassical photon number correlations in tripartite CV systems
thus allowing the simultaneous realization of
ghost-imaging and ghost-diffraction with enhanced sensitivity.
\par\noindent
This work has been supported by MIUR projects PRIN-2005024254-002
and FIRB-RBAU014CLC-002.

\newpage
\end{document}